\documentstyle[epsf]{aa}
\begin{document}
\thesaurus{08 (09.01.1;  09.04.1;   09.18.1) }           
\title{New PAH mode at 16.4 $\mu$m
\thanks{based on observations with ISO, an ESA 
project with instruments funded  by ESA Member States 
(especially the PI countries: the United Kingdom, 
France, the Netherlands, Germany), and with the 
participation of ISAS and NASA.}}
\author{C. Moutou (1), L. Verstraete (2), 
A. L\'eger (2), K. Sellgren (3), W. Schmidt (4)}
\offprints{C. Moutou}
\institute{
(1) ESO, Alonso de Cordoba 3107, Santiago 19, Chile, cmoutou@eso.org\\
(2) IAS, CNRS, Universit\'e Paris Sud, F-91405 Orsay, France, 
verstra@ias.fr, leger@ias.fr \\
(3) Astronomy Dept., Ohio State University, 140 West 18th Avenue,
Columbus OH 43210, USA, sellgren@astronomy.ohio--state.edu\\
(4) Institut f\"ur PAH-Forschung, Flurstrasse 17, 
86926 Greifenberg, Germany\\ }
\date{Received date; accepted date}
\authorrunning{C. Moutou et al.}
\titlerunning{New PAH emission feature at 16.4 $\mu$m}
\maketitle
\begin{abstract}
The detection of a new 16.4 $\mu$m emission feature in the ISO-SWS spectra of 
NGC 7023, M17, and the Orion Bar is reported.
Previous laboratory experiments measured a mode near this wavelength
in spectra of PAHs (Polycyclic Aromatic Hydrocarbons), and so we 
suggest the new interstellar 16.4 $\mu$m
feature could be assigned to low-frequency vibrations of PAHs.
The best carrier candidates seem to be 
PAH molecules containing pentagonal rings.
\end{abstract}
\section{Introduction}
The launch of ISO (Infrared Space Observatory, Kessler et al. 1996) 
has vastly improved our ability to explore
the spectral range beyond 13 $\mu$m, which is blocked by
atmospheric absorption.
Airborne or spaceborne mid-infrared spectrometers prior to 
ISO did not have the sensitivity to determine
whether the carriers 
of the Aromatic Infrared Bands (AIBs), observed in
the interstellar medium (ISM) between 3 and 13 $\mu$m, also emit at
longer wavelengths.

The lack of interstellar data on possible AIBs at 13 -- 25 $\mu$m
meant that laboratory spectroscopy of interstellar
analogs in this spectral range was scanty. 
The hypothesis that polycyclic aromatic hydrocarbons
(PAHs) are responsible for the AIBs
(L\'eger and Puget 1984, Allamandola et al. 1985), 
has led much laboratory work to focus on PAHs.
Spectra longward of 13 $\mu$m   
were initially published for a limited sample 
of small PAH molecules (L\'eger et al. 1989a;
Karcher et al. 1985; Allamandola et al. 1989).
Recently a more extensive laboratory study
(Moutou et al., 1996) has shown that PAH spectra contain 
common vibrational 
frequencies, namely near 16.2, 18.2, 21.3 and 23.2 $\mu$m 
(617, 550, 470 and 430 cm$^{-1}$).
These frequencies where many species exhibit an absorption mode
should correspond to modes with a higher probability of detection in the ISM. 
The IR spectra of many neutral and ionized PAHs isolated in Argon matrices have
also been recently reported (Hudgins and Sandford 1998a,b,c).

Small particles such as PAHs in the ISM
are believed to emit when they are heated to high temperatures 
by transient heating (Sellgren 1984), 
while laboratory absorption spectra are measured at lower temperature.
This results in band broadenings and wavelength shifts of
the interstellar emission features compared to the
laboratory absorption features.
The central wavelengths of PAH vibrational bands
are observed to increase with temperature, and anharmonicity as 
well as mode couplings contribute to the broadening (Joblin et al. 1995). 
We have used the results of Joblin et al. (1995) to correct
the measured central wavelengths of features under 
laboratory (absorption at lower temperature) conditions 
to the expected wavelengths under astronomical (emission
at higher temperature) conditions.

In this paper, we compare the laboratory measurements
of PAHs in the 14 -- 25 $\mu$m domain
to astronomical spectra obtained by ISO.

\section{Observations}
High-resolution Short Wavelength Spectrometer (SWS) spectra
(de Graauw et al. 1996) 
of NGC 7023, M17, and the Orion Bar
were obtained with ISO in a 14$''$$\times$27$''$ beam.
The data reduction was carried out with SWS-IA (update: October 1997).
Bands 3A (12.0 -- 16.5 $\mu$m) and 3C 
(16.5 -- 19.5 $\mu$m) are subject to fringing
(Schaeidt et al. 1996). 
Fringes were removed by subtracting a sine
function, whose amplitude and phase were determined by
least-squares fitting, from the high-frequency part
of each detector scan.
The order edge between bands 3A and 3C
occurs at 16.5 $\mu$m, 
and is visible in Figure \ref{sp}.
A detailed description of the data reduction scheme 
is given in Moutou et al. (1999). 

\subsection{NGC 7023}
\begin{figure}
\vspace{1cm}
\centerline{\epsfxsize=7cm\epsfbox{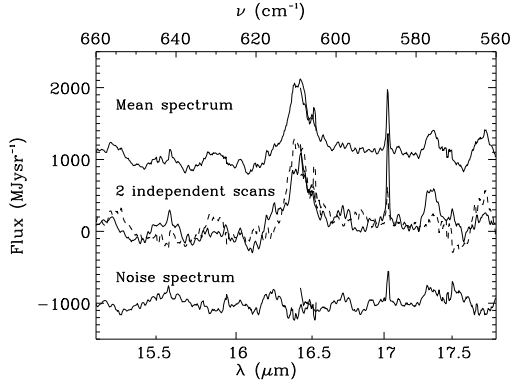}}
\caption{ISO-SWS01 (dashed) and SWS06 (solid) spectra of 
NGC 7023 between 15.1 and 17.8 $\mu$m
(660 -- 560 cm$^{-1}$), at spectral resolution 
$R$ = $\lambda / \Delta \lambda$ = 900 and 1800, respectively.
The two independent scans in the middle 
show the same emission feature at 16.4 $\mu$m,
whereas noise features are not reproduced from one 
spectrum to the other (e.g. at
15.5, 17.4, and 17.6 $\mu$m). 
The 0--0 S(1) pure rotational line of H$_2$ is seen at 17.0 $\mu$m. 
The bottom plot, 
shifted by $-1000$ MJy sr$^{-1}$ for clarity, 
is a noise spectrum obtained by subtracting one spectrum 
from the other after rebinning to $R$ = 900.
The upper spectrum is the average
of both spectra, rebinned to $R$ = 900,
and shifted by $+1000$ MJy sr$^{-1}$.}
\label{sp}
\end{figure}

We observed the bright reflection nebula NGC 7023 
at a position 27$''$ W 34$''$ N of its illuminating star HD 200775. 
The SWS01 spectrum (2.5 -- 45 $\mu$m) 
has a spectral resolution of $R$ = $\lambda / \Delta \lambda$ = 900 
(Sellgren et al. in preparation). Another spectrum
with SWS-AOT6, at $R$ = 1800, provided an independent spectrum in the
reduced domain 14.5 -- 19.5 $\mu$m on which we focus here. 
The continuum level in this object is quite low and 
almost flat between 13 and 25 $\mu$m,
which offers the best conditions for searching for new features.
Comparison of spectra taken at different times is a reliable test 
for distinguishing
noise features from real emission bands.
Both spectra are displayed in Fig. \ref{sp}.
The SWS slit orientation differed between these spectra, 
which explains the varying intensity of the 17.0 $\mu$m H$_2$ line.

\begin{figure}
\centerline{\epsfxsize=6cm\epsfbox{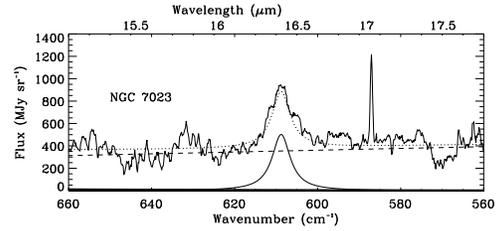}}
\caption{Lorentzian fitting of the 
16.4 $\mu$m band in the SWS01 spectrum of NGC 7023.
The dashed line indicates the underlying continuum and the dotted line shows
the sum of continuum and Lorentzian band. The H$_2$ line was not fitted.}
\label{lor}
\end{figure}

The spectra contain an emission feature at 16.4 $\mu$m,
detected in all independent scans.  This new
emission feature was first mentioned by 
Boulanger et al. (1998) and Tielens et al. (1999).
We have fit the 16.4 $\mu$m band in our SWS data
by a Lorentzian profile (Fig. \ref{lor}). 
Lorentzian fitting of AIB spectra 
has proven to be a useful tool for extracting the
band contribution from the underlying continuum (Boulanger et al. 1998).

\subsection{Detection in other objects}
\begin{figure}
\vspace{1cm}
\centerline{\epsfxsize=7cm\epsfbox{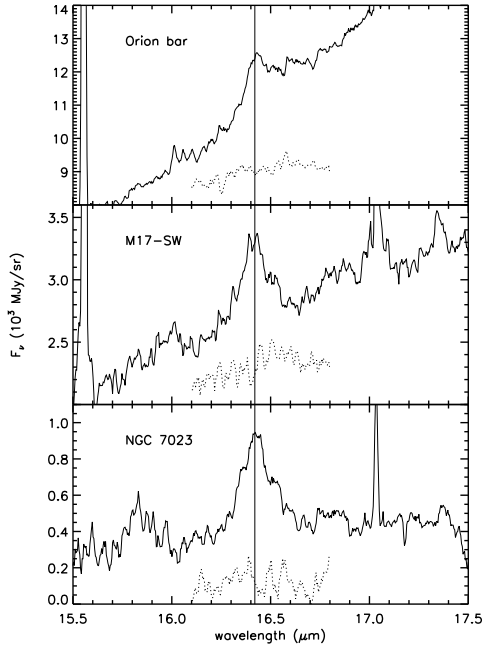}}
\caption{The 16.4 $\mu$m feature in
ISO-SWS01 spectra (solid lines) at $R$ = 900 of the
Orion Bar, M17-SW, and NGC 7023
(see text for positions).
The dotted lines show the difference between
independent upwards and downwards spectra, as a noise estimate.
The strong line at 15.5 $\mu$m in Orion and M17 is [Ne III].}
\label{other}
\end{figure}

As shown in Fig. \ref{other}, the 16.4 $\mu$m band 
is also detected in the ISO-SWS01 ($R$ = 900)
spectra of the Orion Bar 
($\alpha_{2000} = 5:35:20.3$, $\delta_{2000} = -5:25:20$), 
and in M17-SW, at the interface between the HII 
region and the molecular cloud 
($\alpha_{2000} = 18:20:22.1$, $\delta_{2000} = -16:12:41.3$).
More details on these spectra 
are given in Verstraete et al. (in preparation).

Individual values of the band parameters are listed in Table 1. 
The average central wavelength
of the feature is 16.42 $\mu$m (608.9 cm$^{-1}$) 
and the average band width is 0.16 $\mu$m (5.8 cm$^{-1}$). 
The uncertainty in the intensity is based on the difference
between independent scans and does not include the overall
calibration uncertainty of SWS.
Some variation in the band-to-continuum ratio 
at 16.4 $\mu$m is observed, likely due to the stronger radiation field
in M17-SW and the Orion Bar. This will be discussed further in a
forthcoming paper.

\begin{table}[!t]
\begin{flushleft}
\leavevmode
\small
\begin{tabular}[h]{cccc}
\hline
\multicolumn{1}{c}{Parameter}  &  \multicolumn{1}{c}{NGC 7023} &
\multicolumn{1}{c}{M17-SW} & \multicolumn{1}{c}{Orion Bar} \\
\hline
$\nu$ $^1$   &  609.0 &  609.3 &  608.4 \\
FWHM   $^2$      &     6.6 &    5.3 &    5.6 \\
I $^3$     & 1.15 (0.11) & 1.07 (0.17) & 
            2.65 (0.26) \\

\hline
\end{tabular}\\
\caption[]{ Observed parameters of the 16.4 $\mu$m band }
$^1$ Band center in cm$^{-1}$, \\
$^2$ Full width at half maximum in cm$^{-1}$, \\
$^3$ Integrated band intensity in 10$^{-6}$ W m$^{-2}$ sr$^{-1}$. 
The 1$\sigma$ uncertainty is given in parenthesis.
\end{flushleft}
\end{table}
\normalsize

\section{Laboratory spectra}
\label{comp}
The laboratory spectra of many PAHs 
contain a frequently observed mode peaking at 16.2 $\mu$m 
(Moutou et al. 1996; Schmidt et al., not published;
Hudgins \& Sandford 1998a,b,c).
This mode appears slightly shifted from one
molecule to the other, because its frequency depends on the
internal structure of the molecular skeleton.
In total, 29 molecules out of 63 show a band in between 611 and 
623 cm$^{-1}$ (16.21 -- 16.52 $\mu$m). 

The central wavelength of a PAH vibrational mode
should appear redshifted in interstellar spectra, with respect to the position
measured in absorption at cooler temperatures (Joblin et al. 1995).
The averaged redshift observed for coronene over $\sim$300 Kelvins
is approximately 1\% of the vibration wavenumber. 
This would move the measured wavelength of the laboratory 16.2 $\mu$m band 
to the observed wavelength of the interstellar 16.4 $\mu$m band.

Fig. \ref{histo} shows the measured
histograms of the laboratory absorption modes, obtained by counting
the modes, in 5 cm$^{-1}$ bins (or 0.13 $\mu$m bins).
It has been done separately for the seven PAH families
previously defined in the far-infrared data base 
of PAH spectra (Moutou et al. 1996).
This number is a percentage,
normalised to the total number of species in the corresponding family.
Each family contains generally 5 to 9 individual species.
The "PAHs with pentagons" family has 19 components and the "chain PAHs" 
contains 6 additional species taken from the work of Hudgins and Sandford
(1998a,b), leading to a total number of 11 "chain" species.
The 0.16 $\mu$m redshift is applied to the laboratory data.
We deliberately did not take into account in this statistical
approach the relative strength of the modes
and it is thus not similar to a spectrum, because of the
difficulty of combining spectra from three different 
laboratory groups. 
However, the average spectrum of each PAH family measured by Moutou et al.
1996 is displayed in their Figure 1,
and these average laboratory spectra can be directly compared
to the interstellar spectra after applying a 0.16 $\mu$m redshift
to the laboratory data.

We find that the 16.4 $\mu$m band is especially active 
in the spectra of PAHs containing pentagonal rings and of linear PAHs.
In this latest category, the band appears more likely at 16.25$\mu$m and is
dominated by the spectra of small chains containing 2 to 5 rings.
These molecules are not thought to be good candidates as IR emitters, as
they will probably not survive the interstellar radiation fields (Omont, 1986).
Comparatively, the distribution of modes for other kinds of PAHs
in this spectral domain is more spread out in wavelength or has
another accumulation point. 
The contribution
of the PAHs containing pentagons could then possibly dominate
the interstellar emission spectra we observe.

\section{Discussion}
Calculations of infrared spectra are rare for large molecules. 
In our sample of PAHs with pentagons,
only the theoretical spectrum of fluoranthene (C$_{16}$H$_{10}$)
has been calculated (Klaeboe et al. 1981). The
mode is observed at 616 cm$^{-1}$ in the 
laboratory and predicted by simulations to lie at 668 cm$^{-1}$;
it is identified as a B2 in-plane vibration of C--C bonds. It is a complex 
vibration of the global ring structure, which could be described as a
tentative ``rotation'' of the central pentagonal ring 
(J. Brunvoll, private comm.) and consecutive movement of all carbon atoms.
In other molecules, it may also correspond to {\it a global vibration
of the carbon skeleton}. Since the mode is also present in PAHs
without pentagons, it is not clear yet if the five-membered rings have
a special role in the IR activity at 16.4$\mu$m. We investigate in
any case the implications of the presence of pentagonal rings in PAHs:

\begin{figure}
\centerline{\epsfxsize=7.5cm\epsfbox{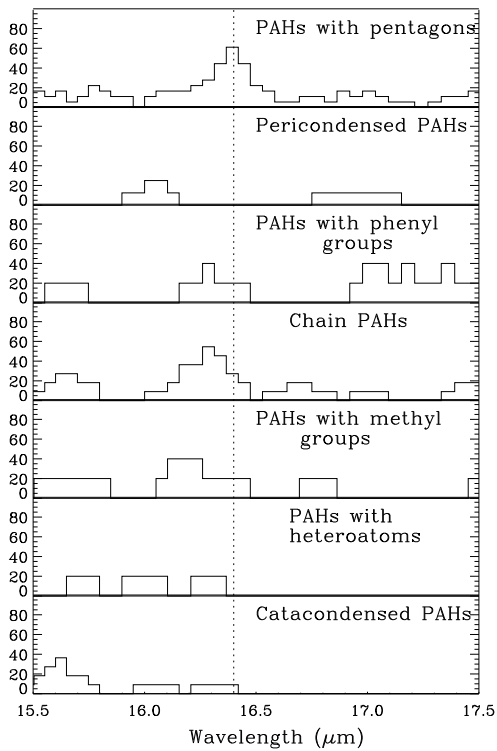}}
\caption{Histograms of PAH modes measured in the laboratory 
(Moutou et al. 1996; Schmidt et al.; Hudgins \& Sandford 1998a,b,c). 
They do not take into account the relative
absorption cross sections. The modes are counted within bins with a 
5 cm$^{-1}$ (0.13 $\mu$m) width while the resolution element of the 
observed spectrum is 0.7 cm$^{-1}$. They are presented as a percentage in 
each family. The dashed line
shows the position of the emission band in the ISO spectrum. A redshift of
6 cm$^{-1}$ (0.16 $\mu$m) has been applied to all laboratory modes, 
in order to take into account the
expected temperature effects ($\approx$ 300 K).}
\label{histo}
\end{figure}

\begin{itemize}
\item{
Pentagonal rings inside a PAH molecule are known to
produce a strong feature around 7 $\mu$m (Moutou et al. 1996, Hudgins \& 
Sandford 1998c).
No individual feature is detected at 7 $\mu$m in NGC 7023,
but we can place an upper limit on 
any 7 $\mu$m band from pentagonal rings from
the 7.0 $\mu$m continuum level (Moutou et al. 1999). 
We adopt the laboratory width of 0.14 $\mu$m for
a pentagonal 7 $\mu$m band. We can then estimate an 
upper limit on the ratio of flux in the 7 $\mu$m band to flux
in the 16.4 $\mu$m band of $<$10 in NGC 7023.
This upper limit on the flux ratio
is not in conflict with the measured laboratory value (2.5).

We conclude that the non-detection of a 7 $\mu$m 
feature would not rule out
a possible signature of pentagons in the ISM at 16.4 $\mu$m .}

\item{
In laboratory spectra where a 16.2 $\mu$m mode is observed, other 
modes may be present, but at positions that vary a
lot from one molecule to another
(Moutou et al. 1996, and Fig. 4). These weak features
are therefore under the detectability limit in our spectra.}

\item{
The fullerene molecule C$_{60}$, which contains pentagonal rings, and its
related cation C$_{60}^+$ are not detected in NGC 7023 (Moutou et al. 1999).
However, the C$_{60}$ infrared spectrum does not show the 16.4 $\mu$m band
(Kr\"atschmer et al. 1990), for symmetry reasons.
The low abundance of 60-atom fullerenes is therefore not 
evidence against pentagonal rings in the ISM.}

\item{
In the
evolution of aromatic molecules, pentagonal rings
tend to curve a molecular plane composed of pure hexagonal PAHs. 
The molecules will then
become tri-dimensional while growing. A global scheme of
aromatic compounds, from simple planar PAHs to small carbon grains or
anthracite coals such as
have been shown to fit the spectra of some planetary nebulae 
(Guillois et al. 1996) requires 
such an intermediate evolution state, where
molecules of $\sim$ 100 carbon atoms are curved. 
It is also consistent with proposals for
grain formation in carbon star shells.
For instance, Goeres \& Seldmayr (1992), with their self-consistent 
model DRACON of carbon chemistry,
predict the formation of large aromatic molecules containing pentagons, 
mostly from C$_3$ accretion, 
because pentagons offer the best trade-off between 
entropy and energy in the process of carbon dust growth. Also,
Kroto \& McKay (1988) proposed an alternative growth sequence to produce carbon
grains, involving pentagonal rings. Their presence in the network would lead to 
``quasi-icosahedral carbon particles with spiral structures'' 
(see also Balm \& Kroto, 1990).}
\end{itemize}

We have shown that the 16.4 $\mu$m mode observed in interstellar spectra could 
be explained by PAH vibrational activity. More work is required to assess the
possible dominant role of five-membered ring PAHs.
Other mid-infrared features should also be searched for in other ISO spectra, 
as the identification of other far-infrared features could give some important 
hints as to the nature of the AIB spectrum.\\[0.3cm]

{\bf Acknowledgments:} We are grateful to J. Brunvoll, S.J. Cyvin and 
P. Klaeboe for very helpful discussions.
Thanks to the anonymous referee for useful comments.


\begin{thebibliography}{}
\bibitem{allam85}
Allamandola, L.J., Tielens, A.G.G.M., \& Barker, J.R., 1985, ApJL 290, L25
\bibitem{allam89}
Allamandola, L.J., Bregman, J.D., Sandford et al., 1989, ApJ, 345, L59
\bibitem{balm}
Balm S.P. \& Kroto H.W., 1990, MNRAS 245, 193
\bibitem{boulanger}
Boulanger F., Boissel P., Cesarsky D. et al., 1998, A\&A 339,194
\bibitem{deg}
de Graauw T., Haser L.N., Beintema D.A. et al., 1996,  A\&A 315, L49
\bibitem{goeres}
Goeres, A. and Seldmayr, E. 1992, A\&A 265, 216
\bibitem{guillois}
Guillois O., Nenner I., Papoular R. et al. 1996, ApJ 464, 810
\bibitem{hudga}
Hudgins D.M. \& Sandford S.A., 1998a, J.Phys Chem. 102, 329
\bibitem{hudgb}
Hudgins D.M. \& Sandford S.A., 1998b, J.Phys Chem. 102, 344
\bibitem{hudgc}
Hudgins  D.M. \& Sandford S.A., 1998c, J.Phys Chem. 102, 353
\bibitem{joblin95}
Joblin C., Boissel P., L\'eger A. et al.,
1995, A\&A 299, 835 
\bibitem{karcher85}
Karcher, W., et al., 1985, 1988, 1991, 
Spectral Atlas of PAH Compounds, Vol 1, Reidel, Dordrecht
and Vols 2\&3, Kluwer, Dordrecht.
\bibitem{ke}
Kessler M., et al. 1996, A\&A 315, L27
\bibitem{klaeboe}
Klaeboe P., Cyvin S.J., Asbojornsen P. et al., 1981, 
Spectrochimica Acta 37A, 655.
\bibitem{kr}
Kr\"atschmer W., Lamb L.D. \& Fostiropoulos K., 1990, Nature 347, 354
\bibitem{kroto}
Kroto H.W \& McKay  K., 1988,  331, 328
\bibitem{lp}
L\'eger, A., Puget, J.L. 1984, A\&A 137, L5 
\bibitem{leger89}
L\'eger A., d'Hendecourt L. \& D\'efourneau D. 1989, A\&A 216, 148
\bibitem{moutb}
Moutou C., L\'eger A., d'Hendecourt L., 1996, A\&A 310, 297
\bibitem{moutc}
Moutou C., Sellgren K., Verstraete L. et al., 1999, A\&A 347, 949
\bibitem{omont}
Omont A., 1986, A\&A 164, 159
\bibitem{schaeidt}
Schaeidt S.G. et al., 1996, A\&A 315, L55
\bibitem{sell84}
Sellgren K., 1984, ApJ 277, 623
\bibitem{tielens}
Tielens A.G.G.M., Hony S., 
van Kerkhoven C. et al., 1999, in ``The Universe as seen by ISO'',
eds P. Cox \& M.F. Kessler, p. 579
\end{thebibliography}
\end{document}